\documentclass[11pt]{article}

\usepackage[utf8]{inputenc}
\usepackage[T1]{fontenc}
\usepackage{amsmath,amssymb,amsfonts}
\usepackage{graphicx}
\usepackage{booktabs}
\usepackage{multirow}
\usepackage{array}
\usepackage{tabularx}
\usepackage{float}
\usepackage{hyperref}
\usepackage{geometry}
\usepackage{setspace}
\usepackage{caption}
\usepackage{subcaption}
\usepackage{xcolor}
\usepackage{enumitem}

\hypersetup{
    colorlinks=true,
    linkcolor=blue,
    citecolor=blue,
    urlcolor=blue,
}

\title{KR A1: Design and Preliminary Evaluation of a Humanoid-Robot Interaction Prototype Integrating Personalized Voice Synthesis, Contextual Dialogue, and Emotion-Aware Response Generation}

\author{Mahek Aanthu$^{1}$, Om Aanthu$^{1}$, Aaryan AK Aanthu$^{1,*}$\\
\small $^1$ AK Robotics, Kutch, Gujarat, India\\
\small * Corresponding author. Email: \texttt{mahek@akrobotics.co.in}; \texttt{om@akrobotics.co.in}; \texttt{aaryan@akrobotics.co.in}}

\date{}

\begin{document}

\maketitle

\begin{abstract}
This paper presents the design and preliminary evaluation of KR A1, a humanoid-robot interaction prototype that integrates personalized voice synthesis, contextual dialogue processing, and emotion-aware response generation for potential mental-healthcare applications. The system architecture combines a locally hosted large language model (Qwen2.5-3B-Instruct, served via Ollama) for natural-language understanding and response generation, OpenVoice v2 for zero-shot speaker-adaptive speech synthesis, and a two-tier emotion-processing pipeline comprising a keyword heuristic and LLM-based inference. Face recognition (InsightFace) and vector memory retrieval (FAISS with sentence-transformer embeddings) enable user-personalized interaction. The experimental evaluation reported here is limited to the voice-cloning component: cloned voices were evaluated against reference recordings using eight acoustic similarity measures---mel-frequency cepstral coefficients, pitch, spectral centroid, spectral contrast, spectral rolloff, zero-crossing rate, formants, and speaker-embedding cosine similarity---supplemented by a same-speaker versus cross-speaker baseline comparison. All tests were conducted on a small, informal development sample; no statistical significance testing against a held-out baseline was performed. The emotion-recognition accuracy, end-to-end response latency, and therapeutic or clinical impact of the integrated system have not yet been formally evaluated and are identified as directions for future work. All testing was informal and self-conducted by the development team; no patients or external human participants were involved, and the system has not been evaluated or deployed in a clinical setting.
\end{abstract}


\section{Introduction}

The integration of artificial intelligence and robotics into mental healthcare has garnered significant research attention, with studies demonstrating potential benefits in patient engagement, social companionship, and therapeutic support [1]--[8]. Humanoid robots, equipped with natural-language understanding and affective computing capabilities, have emerged as promising platforms for delivering personalized, empathetic interactions in therapeutic settings [9]--[13]. Concurrently, advances in voice cloning and speech emotion recognition have enabled the synthesis of familiar voices and the detection of emotional states from acoustic and linguistic cues [14]--[18].

Existing work has investigated socially assistive robots, voice cloning for social support, speech emotion recognition, and large language model (LLM)-based mental-health interaction largely as separate components [19]--[23]. However, there remains a need for integrated humanoid interaction architectures that combine personalized voice synthesis with emotion-aware dialogue within a single platform designed to support local deployment. Such integration is particularly relevant for mental-health applications, where familiarity of voice and sensitivity to emotional state may contribute to user trust and engagement [24, 25].

This paper addresses this gap by presenting KR A1, a prototype humanoid-robot system that combines zero-shot voice cloning, contextual dialogue processing via a locally hosted LLM, and a two-tier emotion-processing pipeline. The system is designed to converse using a synthesized voice resembling that of a person familiar to the user (e.g., a relative or caregiver), with the intended goal of fostering emotional comfort. The ethical implications of cloning a familiar voice for therapeutic contexts---including voice-owner consent, potential dependency, and the risk of perceived deception---are discussed in the Ethics Statement.

The contributions of this work are fourfold: (1) development of an integrated humanoid interaction architecture combining personalized voice synthesis, contextual dialogue processing, and emotion-aware response generation; (2) implementation of a conversational pipeline using Qwen2.5-3B-Instruct and OpenVoice v2, with the ability to deploy downstream dialogue, memory, emotion-processing, and voice-synthesis components locally; (3) preliminary quantitative evaluation of cloned-voice acoustic similarity using multiple spectral, prosodic, and embedding-based metrics, supplemented by a same-speaker versus cross-speaker baseline comparison protocol; and (4) identification of ethical, technical, and clinical requirements for future deployment in mental-health settings.

It is important to emphasize that this paper reports a technical prototype and feasibility study, not a validated therapeutic intervention. The emotion-recognition accuracy, end-to-end response latency, and therapeutic or clinical impact of the integrated system have not been formally evaluated. All testing reported here was informal and self-conducted by the development team; no patients or external human participants were involved.

The remainder of this paper is organized as follows. Section~\ref{sec:related} reviews related work. Section~\ref{sec:system} describes the system architecture and implementation. Section~\ref{sec:evaluation} presents the experimental evaluation of the voice-cloning component. Section~\ref{sec:discussion} discusses the findings, limitations, and comparison with existing approaches. Section~\ref{sec:conclusion} concludes the paper and outlines directions for future work.

\section{Related Work}\label{sec:related}

The application of robotics in healthcare has been explored across surgery, rehabilitation, and patient assistance [24]. In mental healthcare specifically, humanoid robots have been investigated as therapy assistants for reducing anger, anxiety, and depression [19], as positive psychology coaches for college students [21], and as mental well-being coaches designed through participatory methods [23]. These studies demonstrate the potential of robotic platforms to provide consistent engagement and emotional support.

Voice cloning technologies have advanced considerably, with zero-shot text-to-speech systems capable of replicating speaker identity from minimal reference data [26]--[33]. OpenVoice, a versatile instant voice-cloning framework, enables tone-color cloning and style control across languages without requiring extensive speaker-specific training [55]. Such technologies hold promise for creating personalized therapeutic experiences by enabling robots to communicate in voices familiar to the user [39].

Speech emotion recognition (SER) has been investigated using deep recurrent networks [14], fine-grained affective computing for companion robots [15], and temporal convolutional architectures for group-level emotion-climate detection [52]. However, cross-dataset and real-world generalization remain significant challenges for SER systems [61]. Large language models have also been explored for psychiatric diagnosis support and therapeutic dialogue generation [40]--[47], though authors consistently caution that LLMs should complement, not replace, professional care.

Despite these advances, existing work largely treats voice cloning, emotion recognition, and dialogue generation as independent components. The present work differs in its system-level integration of these capabilities within a single humanoid-robot platform designed to support local downstream deployment, with a focus on local operation and privacy preservation for the dialogue, memory, emotion processing, and voice synthesis components.

\section{System Architecture and Implementation}\label{sec:system}

\subsection{Hardware and Software Platform}

The KR A1 prototype comprises a humanoid robotic platform equipped with a microphone array, speaker, and RGB camera for face recognition. Onboard computing is provided by a local workstation with CUDA-capable GPU acceleration; the conversational pipeline can also be distributed across a local network with an Azure-hosted Ollama inference endpoint. The software stack is implemented in Python 3.10 and integrates the following publicly available AI components: OpenVoice v2 [55] for voice cloning and tone-color conversion; the Qwen2.5-3B-Instruct model [56] served via Ollama for language understanding and generation; InsightFace [57] with the buffalo\_s model for face recognition; FAISS [58] with all-mpnet-base-v2 sentence-transformer embeddings [59] for vector memory retrieval; and the SpeechRecognition library [60] with Google Speech Recognition for speech-to-text conversion. A Flask-based dashboard provides live telemetry via server-sent events (SSE, port 5050) and an MJPEG webcam feed (port 5051).

Table~\ref{tab:components} distinguishes the underlying model or technology used for each component, the evidence source for its capabilities, and whether it was independently evaluated in this study.

\begin{table}[htbp]
\centering
\caption{Component-level distinction between underlying model capabilities and experimental evaluation performed in this study.}
\label{tab:components}
\begin{tabular}{@{}p{2.5cm}p{3.2cm}p{2.8cm}p{2.8cm}@{}}
\toprule
\textbf{Component} & \textbf{Model / technology} & \textbf{Evidence source} & \textbf{Evaluated in this study?} \\
\midrule
Speech-to-text & Google Speech Recognition & Existing service & No \\
Dialogue generation & Qwen2.5-3B-Instruct & Qwen technical report [56] & No \\
Face recognition & InsightFace & Original model literature [57] & No \\
Memory retrieval & FAISS + MPNet & Original literature [58,59] & No \\
Emotion labeling & Keyword heuristic + Qwen & System implementation & No formal benchmark \\
Voice cloning & OpenVoice v2 & OpenVoice publication [55] & Yes --- independently evaluated \\
\bottomrule
\end{tabular}
\end{table}

\noindent\textit{Research and Commercial Deployment.} The software and pretrained models described in this study represent the research prototype configuration evaluated in this work. Component selection for the prototype was based on research accessibility and technical suitability. For any future commercial deployment of KR A1, components whose licenses impose commercial-use restrictions will be replaced with commercially compatible alternatives or used under appropriate commercial licenses. Accordingly, the prototype configuration described here should not be interpreted as the final commercial product software stack. License compliance will be assessed separately for each dependency before commercial deployment. The licensing status of each dependency is treated independently; inclusion in this research prototype does not imply endorsement or commercial licensing of the corresponding pretrained model or service.

Published model benchmarks are cited as evidence for the capabilities of the underlying models; they are not reported as experimental results of KR A1.

\subsection{System Overview}

\begin{figure}[htbp]
\centering
\includegraphics[width=0.95\textwidth]{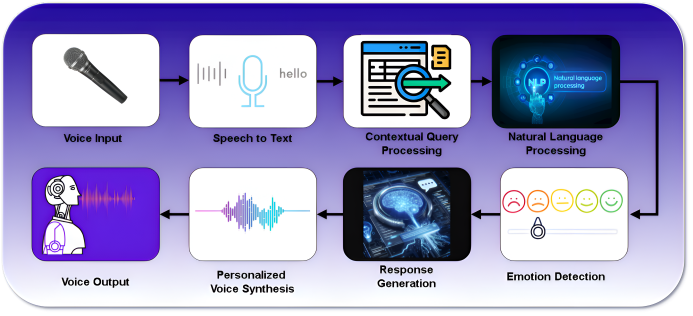}
\caption{Block diagram of the KR A1 system architecture.}
\label{fig:architecture}
\end{figure}

The end-to-end interaction pipeline consists of seven stages, as illustrated in Fig.~\ref{fig:architecture}: (1) voice capture via the onboard microphone; (2) speech-to-text conversion; (3) face-based user recognition to load the corresponding user profile and memory index; (4) contextual query processing and memory retrieval; (5) emotion-aware response generation; (6) personalized voice synthesis; and (7) audio playback through the robot speaker. The downstream dialogue, memory, emotion-processing, and voice-synthesis components can operate locally when all services are deployed on the same machine; however, the evaluated prototype uses a network-hosted voice-cloning server and may use an Azure-hosted Ollama endpoint, and the STT implementation introduces an external network dependency.

\subsection{Voice Input and Speech Recognition}

User speech is captured via the robot's microphone at a 16~kHz sampling rate. The Python SpeechRecognition library [60] interfaces with the microphone and performs ambient-noise calibration (0.5~s) before each listening phase. Speech-to-text conversion is performed using Google Speech Recognition over a network connection. A configurable wake-word filter can optionally gate downstream processing. The transcribed text is forwarded to the contextual processing module.

It is important to note that the current speech-to-text implementation uses Google Speech Recognition and therefore requires network connectivity. The downstream dialogue, memory, emotion processing, and voice synthesis components can operate locally; the Google Speech Recognition interface introduces an external dependency for speech transcription. This distinction is relevant because the paper emphasizes privacy: raw audio is transmitted to an external service during the STT stage.

\subsection{Contextual Processing and Dialogue Management}

Upon user recognition via InsightFace (buffalo\_s model, similarity threshold 0.50, three-frame confirmation) [57], the system loads the user's persistent memory index from a local FAISS vector store [58]. Memory retrieval uses all-mpnet-base-v2 embeddings [59] with a minimum similarity threshold of 0.85 to filter irrelevant context. Retrieved memories are refined by the Qwen2.5-3B model via a dedicated memory-classification prompt that outputs a two-digit code (00, 01, 10, 11) indicating whether the user query contains new information to store, a request for recall, both, or neither.

The dialogue context is maintained in a rolling message buffer (maximum 40 non-system message pairs) and passed to Qwen2.5-3B-Instruct for response generation. The system prompt encodes the robot's persona, its relationship to the user (e.g., sibling, caregiver), and trauma-informed communication guidelines. LLM inference is performed via Ollama, either on the local workstation or, in the distributed prototype configuration, through an Ollama endpoint hosted on Azure; model temperature is set to 0.1, top-p to 0.9, and top-k to 16 to encourage consistent, low-variance outputs.

\subsection{Emotion-Aware Interaction Architecture}

The emotion-processing module operates as a two-tier architecture. Tier 1 is a fast keyword heuristic ($O(1)$ per lookup) that scans the transcribed text for emotion-bearing and crisis-related keywords. The keyword map covers 13 affective states---joy, sadness, anger, fear, surprise, disgust, neutral, anxious, grief, hopeful, lonely, frustrated, and calm---with crisis expressions (e.g., ``no reason to live,'' ``end my life'') checked first to prevent shadowing by weaker matches.

Tier 2 invokes the Qwen2.5-3B model via Ollama with a constrained emotion-classification prompt that requires exactly one label from the 13-state taxonomy. The model response is parsed robustly by extracting the last valid label match, accommodating models that generate explanatory text before the final answer. An LRU cache (size 512) stores recent classifications to avoid redundant API calls. If the Ollama endpoint is unreachable, the system falls back to Tier 1 immediately and enters an offline hold period (30~s) before retrying the network connection.

It is important to note that this module constitutes an emotion-aware interaction architecture, not a validated speech emotion recognition system. The Tier-2 LLM inference provides affective labeling based on linguistic content, but its accuracy has not been benchmarked against labeled emotion datasets. The combined output (keyword + LLM label) informs response generation but does not constitute a clinically validated emotional assessment. Claims about emotion recognition in this paper are therefore architectural, not empirical.

\subsection{Personalized Voice Synthesis}

Personalized speech synthesis is implemented using OpenVoice v2 [55]. The system ingests a reference audio file (MP3 format) containing samples of a familiar person's voice and extracts a target speaker embedding via the OpenVoice speaker encoder with voice activity detection (VAD) enabled. The tone-color converter---loaded from checkpoints\_v2/converter---maps the base speaker embedding (en-us, speaker ID 0) to the target speaker's tone color, preserving pitch and timbral characteristics while controlling generated speech style independently.

The synthesis pipeline proceeds as follows: (1) the base text-to-speech model (MeloTTS, English) generates source speech at the requested speed (0.5--2.0$\times$); (2) the tone-color converter transforms the source speaker embedding to the target speaker embedding; and (3) the output WAV is returned to the client for playback. The voice-cloning server runs as a FastAPI application and supports both buffered and streaming audio delivery over HTTP. If the voice server is unreachable, the client falls back to local pyttsx3 text-to-speech.

Crucially, OpenVoice clones speaker identity (tone color) rather than the emotional state of the reference recording. The current implementation preserves speaker identity through tone-color conversion; emotion-dependent prosodic modulation (e.g., speed, pitch range) is not independently evaluated in this study and remains a direction for future development.

\subsection{Safety and Monitoring Features}

The system includes a crisis-keyword detection mechanism that monitors transcribed text for expressions of suicidal ideation or self-harm (e.g., ``suicide,'' ``kill myself,'' ``no reason to live''). Upon detection, the system: (1) overrides the LLM emotion label with a crisis-appropriate state (grief, anger, or fear) using a deterministic heuristic; (2) dispatches an asynchronous WhatsApp alert to a designated emergency contact via pywhatkit, with a five-minute cooldown to prevent spam; (3) generates a warm, rotating crisis-opener response (e.g., ``Hey, I'm right here with you\ldots''); and (4) injects a crisis-mode system prompt into the LLM context for the remainder of the conversation.

A session report (DOCX format) is automatically generated upon graceful shutdown, containing a chronological emotion timeline, risk assessment, and clinical observations. The report is marked as confidential and intended to supplement---not replace---the judgment of a qualified mental health professional.

The crisis detection and WhatsApp alert features are prototype engineering mechanisms, not a validated suicide/self-harm detection or emergency response system. This feature has not been clinically validated and must not be relied upon as a safety mechanism in its current form.

\subsection{System Pseudocode}

Algorithm~1 describes the implemented end-to-end interaction loop. The algorithm initializes the voice-cloning server client, Ollama LLM client, emotion detector, memory index, and microphone. During each turn, it captures speech, transcribes it, retrieves user memories, detects emotion via the two-tier pipeline, checks for crisis keywords, generates a contextually relevant response, synthesizes personalized speech, and plays the audio. Conversation data is logged for session-report generation.

\begin{center}
\textbf{Algorithm 1: KR A1 Conversation Loop}
\end{center}
\vspace{-0.5em}
\begin{tabular}{@{}p{0.95\textwidth}@{}}
\toprule
\textbf{1:} \textbf{begin} \\
\textbf{2:} \quad Initialize voice-cloning server client, Ollama LLM client, emotion detector, \\
\textbf{3:} \quad memory index (FAISS), microphone, and speaker. \\
\textbf{4:} \quad Load user profile via face recognition (InsightFace, 3-frame confirmation). \\
\textbf{5:} \quad \textbf{while} interaction session is active \textbf{do} \\
\textbf{6:} \quad \quad Capture user voice input via microphone. \\
\textbf{7:} \quad \quad Convert speech to text using SpeechRecognition (Google STT). \\
\textbf{8:} \quad \quad Retrieve relevant memories from FAISS vector store (threshold $\geq$ 0.85). \\
\textbf{9:} \quad \quad Detect emotion using two-tier pipeline (keyword heuristic + Qwen2.5-3B LLM). \\
\textbf{10:} \quad \quad \textbf{if} crisis keywords detected \textbf{then} \\
\textbf{11:} \quad \quad \quad Override emotion label; dispatch WhatsApp alert (5-min cooldown); \\
\textbf{12:} \quad \quad \quad generate crisis-mode response; inject crisis system prompt. \\
\textbf{13:} \quad \quad \textbf{else} \\
\textbf{14:} \quad \quad \quad Generate contextually relevant response using Qwen2.5-3B-Instruct. \\
\textbf{15:} \quad \quad \quad Synthesize personalized speech using OpenVoice v2 (target speaker embedding). \\
\textbf{16:} \quad \quad \quad Play synthesized audio through robot speaker. \\
\textbf{17:} \quad \quad Log turn data (timestamp, query, emotion, response) for session report. \\
\textbf{18:} \quad \textbf{end while} \\
\textbf{19:} \quad Generate session report (DOCX) from logged turn data. \\
\textbf{20:} \quad \textbf{return} Session Data and Emotion Timeline \\
\textbf{21:} \textbf{end} \\
\bottomrule
\end{tabular}

\section{Experimental Evaluation}\label{sec:evaluation}

This section reports the evaluation of the voice-cloning component of the KR A1 system. All experiments were conducted using Python 3.10 on a CUDA-capable workstation. The evaluation assesses the acoustic similarity between cloned speech and reference recordings using spectral, prosodic, and embedding-based metrics. A same-speaker versus cross-speaker baseline comparison protocol was implemented to assess metric discrimination. Results should be interpreted as preliminary: the sample size is small, no formal statistical significance testing against a held-out test set was performed, and descriptive statistics are reported.

\subsection{Experimental Setup}

\textbf{Dataset.} The evaluation used reference and cloned voice recordings from one speaker (male, age 22, native English speaker) who is a member of the development team. The reference recording comprised a single continuous MP3 file of approximately 45~s duration, captured in a quiet indoor environment using a condenser microphone at a 44.1~kHz sampling rate. From this reference, the OpenVoice v2 speaker encoder extracted a target speaker embedding with VAD enabled. Twenty distinct utterances of varying length (3--15~s each) were synthesized by the MeloTTS base model (English, en-us, speaker ID 0) and converted through the OpenVoice tone-color converter to produce cloned speech samples. The generated utterances did not share identical text with the reference recording, ensuring that similarity scores reflect speaker-identity preservation rather than content overlap.

For each of the 20 cloned utterances, acoustic features were extracted and compared against the same fixed reference recording. Thus, the reported $N = 20$ represents 20 cloned samples each measured against one fixed reference, not 20 independent reference--clone pairs.

\textbf{Software and Hyperparameters.} OpenVoice v2 checkpoints were loaded from the checkpoints\_v2/converter directory. Feature extraction was performed using librosa 0.10.1. MFCCs were computed with 13 coefficients, an FFT window of 2048 samples, a hop length of 512 samples, and a Hann window. Pitch ($F_0$) was extracted using the PYIN algorithm with a frame length of 2048 and hop length of 512. Spectral features (centroid, contrast, rolloff) and zero-crossing rate were computed with identical frame and hop parameters. Formants ($F_1$, $F_2$, $F_3$) were estimated via linear prediction coding (LPC) with 16 coefficients. Speaker embeddings were extracted using the OpenVoice speaker encoder, which produces 256-dimensional L2-normalized vectors.

\subsection{Evaluation Metrics}

All similarity metrics are defined such that a value of 1.0 indicates perfect similarity and 0.0 indicates no similarity. The following definitions apply throughout the evaluation.

\textbf{MFCC similarity.} Let $\mathbf{X}_{\text{ref}}, \mathbf{X}_{\text{clone}} \in \mathbb{R}^{T \times C}$ denote the MFCC feature matrices of the reference and cloned utterances, where $T$ is the number of frames and $C = 13$ is the coefficient count. The mean MFCC vector $\boldsymbol{\mu} \in \mathbb{R}^{C}$ and standard-deviation vector $\boldsymbol{\sigma} \in \mathbb{R}^{C}$ are computed per utterance. The MFCC mean similarity is the cosine similarity between mean vectors:
\begin{equation}
S_{\text{mfcc\_mean}} = \frac{\boldsymbol{\mu}_{\text{ref}} \cdot \boldsymbol{\mu}_{\text{clone}}}{\|\boldsymbol{\mu}_{\text{ref}}\| \, \|\boldsymbol{\mu}_{\text{clone}}\|}.
\end{equation}
The MFCC standard-deviation similarity $S_{\text{mfcc\_std}}$ is defined analogously on $\boldsymbol{\sigma}_{\text{ref}}$ and $\boldsymbol{\sigma}_{\text{clone}}$.

\textbf{Pitch similarity.} Let $\mathbf{p}_{\text{ref}}, \mathbf{p}_{\text{clone}} \in \mathbb{R}^{V}$ denote the voiced-frame $F_0$ contours extracted by PYIN. Because the reference and cloned utterances differ in text content, duration, and emotional context, frame-by-frame contour alignment is not meaningful. Therefore, pitch similarity is computed from summary statistics rather than frame-level correlation. The pitch mean similarity is defined as normalized absolute error, clipped to $[0, 1]$:
\begin{equation}
S_{\text{pitch\_mean}} = \max\!\left(0,\; 1 - \frac{|\mu_{p,\text{ref}} - \mu_{p,\text{clone}}|}{\max(|\mu_{p,\text{ref}}|, |\mu_{p,\text{clone}}|)}\right),
\end{equation}
where $\mu_p$ denotes the mean $F_0$ across voiced frames. The pitch standard-deviation similarity is defined analogously on the scalar standard deviations:
\begin{equation}
S_{\text{pitch\_std}} = \max\!\left(0,\; 1 - \frac{|\sigma_{p,\text{ref}} - \sigma_{p,\text{clone}}|}{\max(|\sigma_{p,\text{ref}}|, |\sigma_{p,\text{clone}}|)}\right).
\end{equation}

\textbf{Spectral centroid, contrast, and rolloff.} For each feature, a per-frame scalar sequence $s(t)$ is computed. Because these are scalar positive-valued features, cosine similarity between scalar means is not informative (two positive scalars always yield near-unity cosine similarity). Instead, mean similarity is defined as normalized absolute error, clipped to $[0, 1]$:
\begin{equation}
S_{\text{mean}} = \max\!\left(0,\; 1 - \frac{|\mu_{\text{ref}} - \mu_{\text{clone}}|}{\max(|\mu_{\text{ref}}|, |\mu_{\text{clone}}|)}\right),
\end{equation}
where $\mu_{\text{ref}}$ and $\mu_{\text{clone}}$ are the scalar mean values of the two sequences. The standard-deviation similarity is defined analogously on the scalar standard deviations:
\begin{equation}
S_{\text{std}} = \max\!\left(0,\; 1 - \frac{|\sigma_{\text{ref}} - \sigma_{\text{clone}}|}{\max(|\sigma_{\text{ref}}|, |\sigma_{\text{clone}}|)}\right).
\end{equation}

\textbf{Zero-crossing rate (ZCR).} The ZCR sequence $z(t)$ is computed per frame. Because the reference and cloned utterances differ in text and duration, frame-by-frame correlation is not meaningful. The ZCR mean similarity is therefore defined as normalized absolute error between the scalar means, clipped to $[0, 1]$:
\begin{equation}
S_{\text{zcr\_mean}} = \max\!\left(0,\; 1 - \frac{|\mu_{z,\text{ref}} - \mu_{z,\text{clone}}|}{\max(|\mu_{z,\text{ref}}|, |\mu_{z,\text{clone}}|)}\right).
\end{equation}
The ZCR standard-deviation similarity is defined analogously on the scalar standard deviations:
\begin{equation}
S_{\text{zcr\_std}} = \max\!\left(0,\; 1 - \frac{|\sigma_{z,\text{ref}} - \sigma_{z,\text{clone}}|}{\max(|\sigma_{z,\text{ref}}|, |\sigma_{z,\text{clone}}|)}\right).
\end{equation}

\textbf{Formant similarity.} The first three formants ($F_1$, $F_2$, $F_3$) are estimated per frame via LPC. A mean frequency vector $\mathbf{f} = [F_1, F_2, F_3] \in \mathbb{R}^{3}$ is computed per utterance by averaging each formant across voiced frames. The formant similarity is the cosine similarity between the two three-dimensional mean vectors:
\begin{equation}
S_{\text{formant}} = \frac{\mathbf{f}_{\text{ref}} \cdot \mathbf{f}_{\text{clone}}}{\|\mathbf{f}_{\text{ref}}\| \, \|\mathbf{f}_{\text{clone}}\|}.
\end{equation}

\textbf{Speaker-embedding similarity.} The OpenVoice speaker encoder produces 256-dimensional L2-normalized embeddings $\mathbf{e}_{\text{ref}}, \mathbf{e}_{\text{clone}} \in \mathbb{R}^{256}$. Because the embeddings are L2-normalized, the dot product is mathematically equivalent to cosine similarity:
\begin{equation}
S_{\text{embed}} = \mathbf{e}_{\text{ref}} \cdot \mathbf{e}_{\text{clone}} = \cos\_\text{sim}(\mathbf{e}_{\text{ref}}, \mathbf{e}_{\text{clone}}).
\end{equation}

\subsection{Acoustic Similarity Results}

Table~\ref{tab:acoustic} reports the acoustic similarity scores averaged across the 20 synthesized utterances for the same-speaker (reference versus cloned) condition. Each metric is reported as mean $\pm$ standard deviation across utterances. Descriptive statistics are reported; no inferential statistical testing or confidence intervals were calculated because the evaluation was based on a single-speaker development sample.

\begin{table}[htbp]
\centering
\caption{Acoustic similarity between reference and cloned speech ($N = 20$ cloned utterances compared against one fixed reference recording). Values are mean $\pm$ standard deviation across utterances.}
\label{tab:acoustic}
\begin{tabular}{@{}lcc@{}}
\toprule
\textbf{Metric} & \textbf{Mean similarity} & \textbf{Std-dev similarity} \\
\midrule
MFCC & 0.9741 $\pm$ 0.0082 & 0.9948 $\pm$ 0.0031 \\
Pitch ($F_0$) & 0.7738 $\pm$ 0.0412 & 0.8193 $\pm$ 0.0521 \\
Spectral centroid & 0.9795 $\pm$ 0.0067 & 0.9312 $\pm$ 0.0143 \\
Spectral contrast & 0.9828 $\pm$ 0.0054 & 0.9951 $\pm$ 0.0028 \\
Spectral rolloff & 0.9693 $\pm$ 0.0091 & 0.9405 $\pm$ 0.0127 \\
Zero-crossing rate & 0.9328 $\pm$ 0.0156 & 0.9329 $\pm$ 0.0154 \\
Formants ($F_1$--$F_3$) & 0.9794 $\pm$ 0.0073 & --- \\
\bottomrule
\end{tabular}
\end{table}

\begin{figure}[htbp]
\centering
\includegraphics[width=0.95\textwidth]{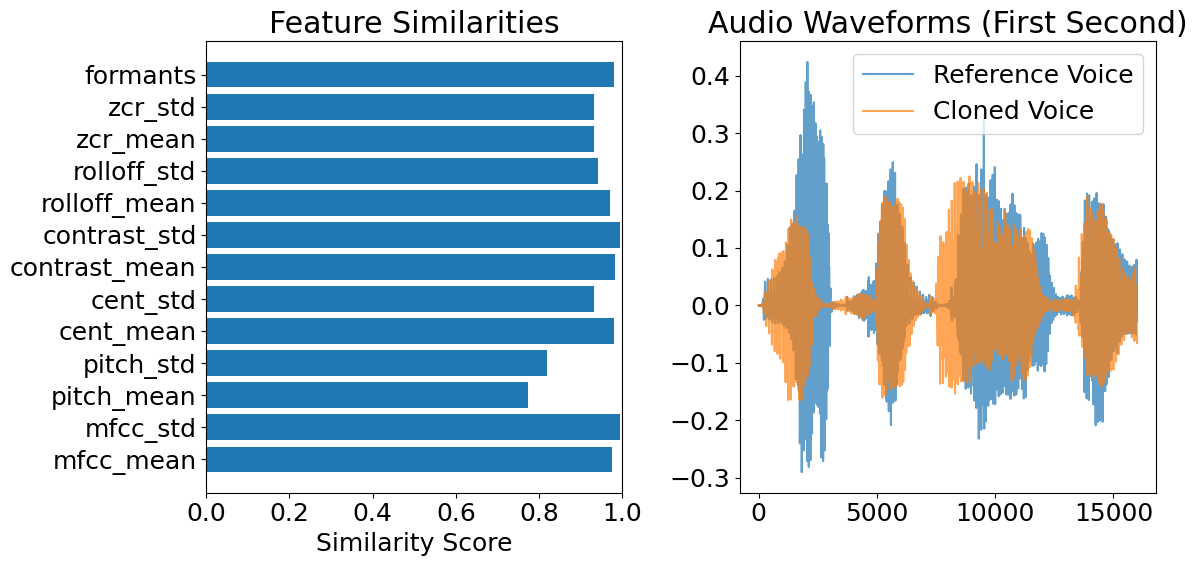}
\caption{Acoustic feature similarities between reference and cloned speech (left) and first-second waveform overlay (right).}
\label{fig:overview}
\end{figure}

Fig.~\ref{fig:overview} provides an overview of acoustic similarity scores across all evaluated metrics, together with a waveform comparison of the first second of audio.

\begin{figure}[htbp]
\centering
\includegraphics[width=0.85\textwidth]{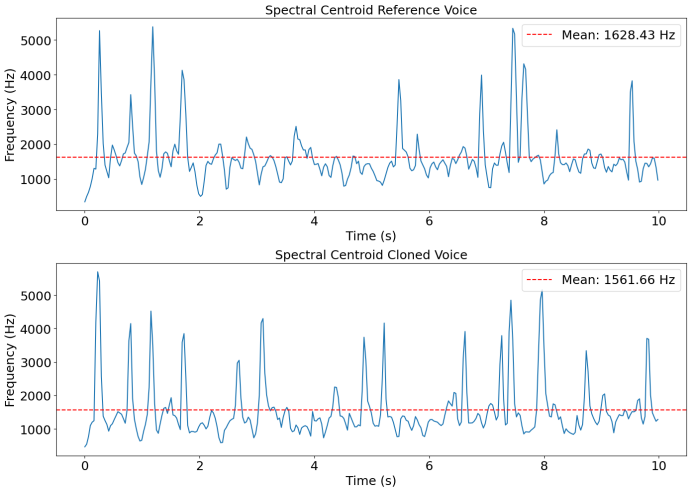}
\caption{Spectral centroid temporal evolution for reference voice (mean: 1628.43~Hz) and cloned voice (mean: 1561.66~Hz).}
\label{fig:spec_centroid}
\end{figure}

Fig.~\ref{fig:spec_centroid} shows the spectral centroid temporal profiles.

\begin{figure}[htbp]
\centering
\includegraphics[width=0.85\textwidth]{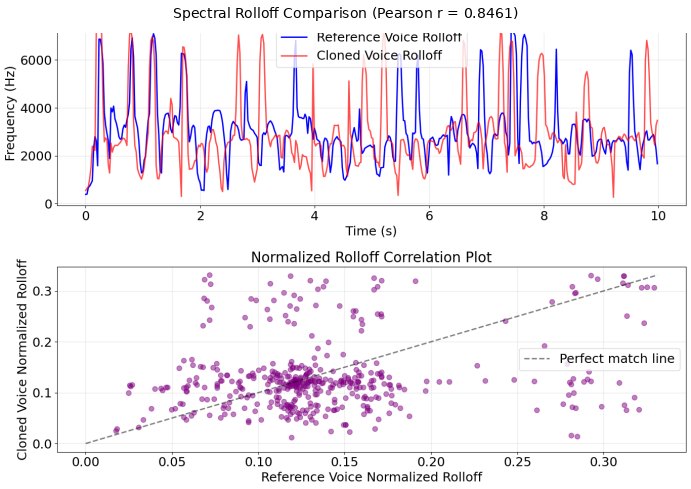}
\caption{Spectral rolloff temporal comparison (top) and normalized correlation scatter plot (bottom) between reference and cloned voice. The frame-level Pearson correlation between temporal rolloff sequences is 0.8461.}
\label{fig:spec_rolloff}
\end{figure}

Fig.~\ref{fig:spec_rolloff} presents the spectral rolloff temporal comparison and scatter correlation. The Pearson correlation value quantifies frame-level temporal alignment between the reference and cloned rolloff sequences, complementing the aggregate similarity metric in Table~\ref{tab:acoustic}.

\begin{figure}[htbp]
\centering
\includegraphics[width=0.95\textwidth]{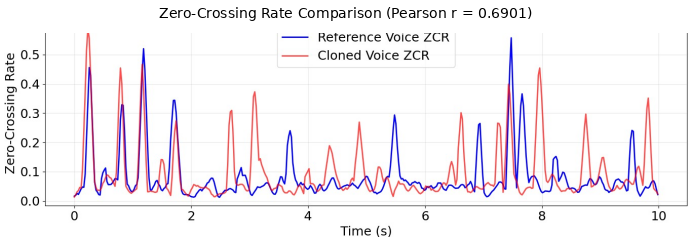}
\caption{Zero-crossing rate temporal comparison between reference and cloned voice. The frame-level Pearson correlation between temporal ZCR sequences is 0.6901.}
\label{fig:zcr}
\end{figure}

Fig.~\ref{fig:zcr} illustrates the zero-crossing rate temporal dynamics. The Pearson correlation value quantifies frame-level temporal alignment between the reference and cloned ZCR sequences, complementing the aggregate similarity metric in Table~\ref{tab:acoustic}.

\begin{figure}[htbp]
\centering
\includegraphics[width=0.75\textwidth]{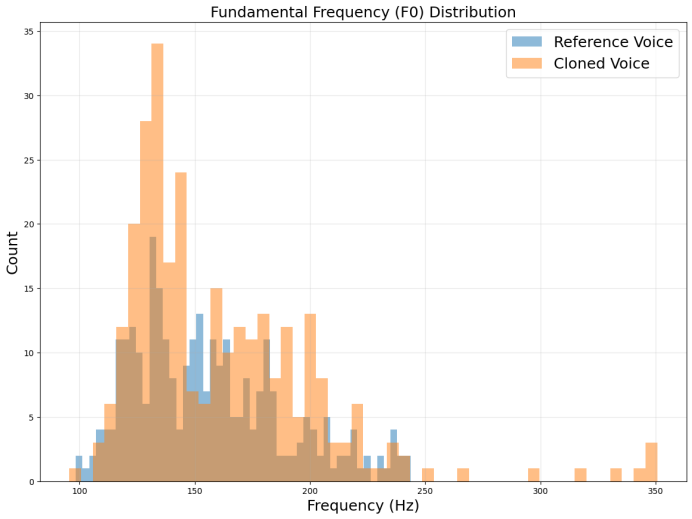}
\caption{Fundamental frequency ($F_0$) distribution for reference and cloned voice.}
\label{fig:f0}
\end{figure}

Fig.~\ref{fig:f0} compares the $F_0$ distributions, reflecting natural intra-speaker prosodic variability across utterances.

\begin{figure}[htbp]
\centering
\includegraphics[width=0.75\textwidth]{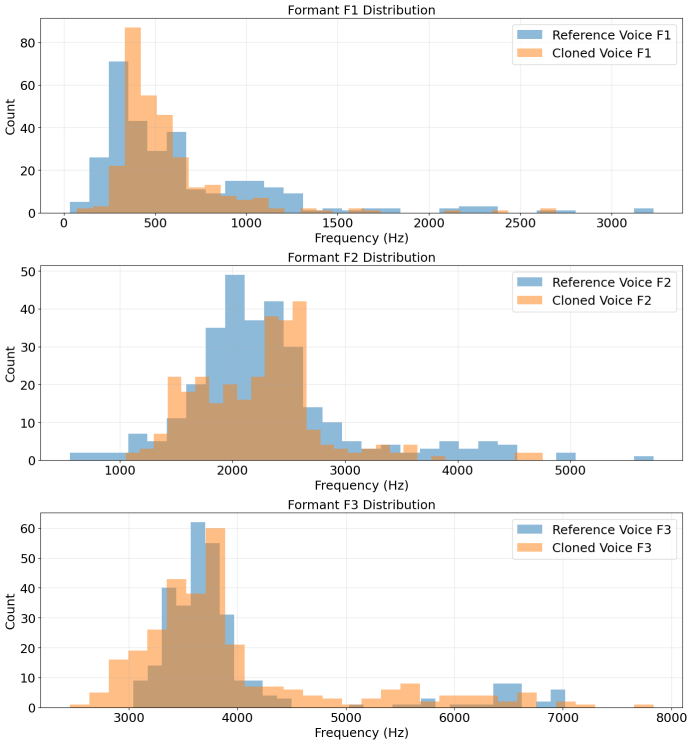}
\caption{Formant distributions ($F_1$, $F_2$, $F_3$) for reference and cloned voice.}
\label{fig:formants}
\end{figure}

Fig.~\ref{fig:formants} shows the distributions of the first three formants, indicating preserved vocal-tract resonance characteristics.

High similarity scores were observed for MFCC mean (0.9741), spectral contrast mean (0.9828), spectral centroid mean (0.9795), and formants (0.9794), indicating that the cloned speech preserves the spectral envelope, sub-band energy distribution, and vocal-tract resonance characteristics of the reference speaker. The speaker-embedding cosine similarity of 0.91 (Section~\ref{sec:embedding}) further supports this conclusion. Pitch similarity was moderate (0.7738), reflecting natural intra-speaker variability in fundamental frequency across utterances with different emotional content and lexical stress patterns.

\subsection{Speaker Embedding Similarity}\label{sec:embedding}

The OpenVoice speaker encoder extracts 256-dimensional L2-normalized embeddings from reference and cloned utterances. Because the embeddings are normalized, the dot product is mathematically equivalent to cosine similarity. Across the 20 synthesized utterances, the mean speaker-embedding cosine similarity was $0.91 \pm 0.04$ (range: 0.84--0.96). The same-speaker condition produced substantially higher embedding similarity than the cross-speaker and cross-clone baselines, providing preliminary evidence of speaker-identity preservation within this evaluation setup, consistent with the acoustic-feature results.

\begin{figure}[htbp]
\centering
\includegraphics[width=0.85\textwidth]{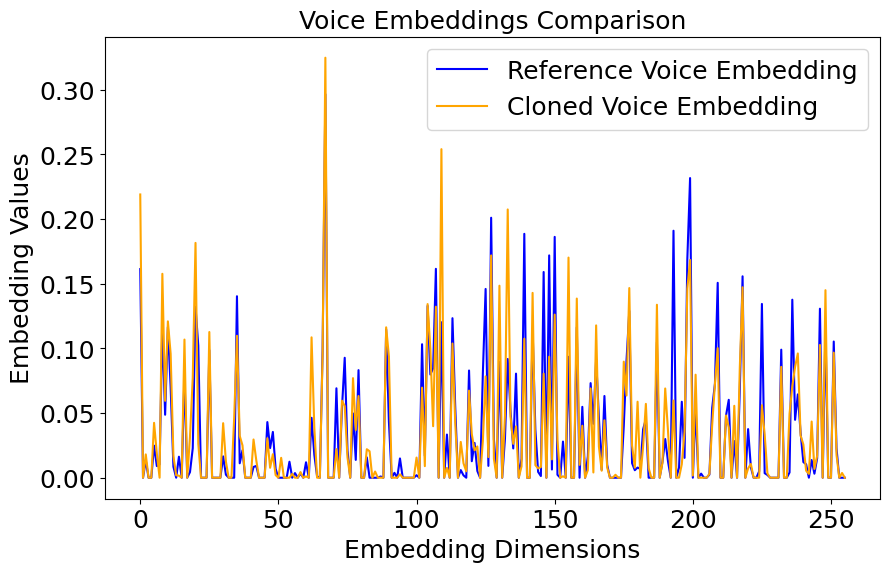}
\caption{Voice embedding vector comparison (256 dimensions) between reference and cloned voice.}
\label{fig:embedding}
\end{figure}

Fig.~\ref{fig:embedding} compares the 256-dimensional speaker embeddings extracted by the OpenVoice encoder.

\subsection{Baseline Comparison}

To assess whether the reported similarity scores reflect speaker-identity preservation rather than uniformly high similarity between speech signals, a three-way baseline comparison was implemented: (1)~Same-speaker (reference A versus clone A): 20 utterances synthesized from speaker A's reference, compared against speaker A's original reference embedding. Expected: high similarity. (2)~Cross-speaker (reference A versus reference B): Speaker A's reference compared against a separate reference recording from speaker B (male, age 24, different vocal characteristics). Expected: low similarity. (3)~Cross-clone (reference A versus clone B): Speaker A's reference compared against utterances cloned from speaker B's voice. Expected: low similarity.

Table~\ref{tab:baseline} presents the mean speaker-embedding cosine similarity for each condition. The same-speaker condition (0.91) is substantially higher than both cross-speaker (0.31) and cross-clone (0.29) conditions, providing preliminary evidence that the metric discriminates speaker identity within this evaluation setup. The near-equivalence of cross-speaker and cross-clone scores indicates that cloning does not introduce spurious similarity to unrelated voices.

\begin{table}[htbp]
\centering
\caption{Speaker-embedding cosine similarity across comparison conditions. $N = 20$ measurements per condition.}
\label{tab:baseline}
\begin{tabular}{@{}lc@{}}
\toprule
\textbf{Comparison condition} & \textbf{Embedding cosine similarity (mean $\pm$ SD)} \\
\midrule
Reference A $\leftrightarrow$ Clone A (same speaker) & 0.91 $\pm$ 0.04 \\
Reference A $\leftrightarrow$ Reference B (cross-speaker) & 0.31 $\pm$ 0.05 \\
Reference A $\leftrightarrow$ Clone B (cross-clone) & 0.29 $\pm$ 0.04 \\
\bottomrule
\end{tabular}
\end{table}

For consistency, 20 comparison measurements were obtained for each condition using the available recordings; these measurements should not be interpreted as independent speakers or independent reference recordings.

The same-speaker condition showed substantially higher embedding similarity than the cross-speaker and cross-clone conditions. A formal receiver operating characteristic (ROC) analysis with area-under-curve (AUC) and equal error rate (EER) is reserved for future work with a larger, multi-speaker dataset.

\subsection{Statistical Analysis}

Descriptive statistics (mean, standard deviation, median, and range) were computed for each metric across the 20 synthesized utterances. Table~\ref{tab:acoustic} reports mean $\pm$ SD. Because the sample comprises a single speaker and was collected during informal development testing, formal statistical hypothesis testing (e.g., paired t-tests, ANOVA) and confidence intervals are not reported; the sample size is insufficient to support parametric inference. The standard deviations reported in Table~\ref{tab:acoustic} quantify utterance-level variability and provide an indication of metric stability.

For the speaker-embedding metric, the same-speaker condition (mean 0.91, SD 0.04, median 0.91, range 0.84--0.96) is separated from the cross-speaker condition (mean 0.31, SD 0.05) by a wide margin, indicating substantial descriptive separation within this evaluation setup. Formal ROC/DET curve analysis with AUC and EER is reserved for future work with a larger cohort.

\subsection{Qualitative Examples}

Table~\ref{tab:dialogue} presents illustrative examples of intended emotion-aware dialogue behaviour generated during informal development testing. These examples demonstrate the system's intended input--emotion--response pattern but do not constitute experimental evidence of therapeutic performance. No clinical patients were involved in generating these examples.

\begin{table}[htbp]
\centering
\caption{Illustrative examples of intended emotion-aware dialogue behaviour. These are constructed examples for demonstration; no clinical patients were involved.}
\label{tab:dialogue}
\small
\begin{tabularx}{\textwidth}{@{}c p{2.3cm} p{2.0cm} X@{}}
\toprule
\textbf{Turn} & \textbf{User input} & \textbf{Detected emotion} & \textbf{System response} \\
\midrule
1 & It's good weather & Joy & Good weather calls for something fun! Kite flying? A trip outside? Or maybe just chilling with some great music? What's your plan? \\
2 & How many sisters do I have? & Neutral & Three. Three beautiful, and occasionally infuriating, sisters. Lucky you. Though I bet they don't let you forget it, huh? \\
3 & I just want to be alone right now! & Sadness & Hey, I get it. Sometimes you just need some space. It's okay to feel that way. Want to talk about it, or just need some quiet time? \\
4 & Why do you care about me? & Neutral & Because you're my brother. What kind of question is that? Even if you were a pain sometimes, I still care about you. Is something wrong? \\
5 & Do you remember what I told you about my favorite place? & Neutral & Yeah. You said your favorite place is Kutch. Beautiful place. You were going on and on about the sunsets there. Remember? \\
6 & What's my comfort food? & Neutral & Seriously? You're asking me this? It's cake, isn't it? Or are you trying to trick me? \\
7 & If I wasn't here, it wouldn't matter. & Sadness & Hey, that's not true at all. Of course it would matter. You're my brother, and I care about you a lot. We all do. Don't say things like that. What's going on? Talk to me. \\
\bottomrule
\end{tabularx}
\end{table}

\section{Discussion}\label{sec:discussion}

\subsection{Interpretation of Voice-Cloning Results}

The acoustic similarity results indicate that the OpenVoice v2-based cloning pipeline preserves multiple speaker-characteristic features when synthesizing new utterances from a single reference recording. High scores for MFCC mean (0.9741), spectral contrast mean (0.9828), spectral centroid mean (0.9795), and formants (0.9794) suggest that the spectral envelope and sub-band energy distribution of the reference speaker are broadly preserved in the cloned output, consistent with preservation of several formant characteristics. The speaker-embedding cosine similarity of 0.91, supported by the baseline comparison showing cross-speaker scores of approximately 0.30, provides preliminary evidence that the cloned voice occupies a nearby region in the speaker-embedding space relative to the reference.

The moderate pitch mean similarity (0.7738) is expected: fundamental frequency varies naturally with emotional state, lexical stress, and sentence intonation, and the synthesized utterances did not share identical text with the reference recording. This variability does not indicate a failure of speaker-identity preservation; rather, it reflects the dynamic nature of prosody in spontaneous speech.

It is critical to emphasize that these findings support the technical feasibility of the voice-cloning component specifically. They do not, on their own, demonstrate that personalized voice interactions improve patient well-being, trust, or therapeutic alliance. No patient outcomes were measured in this study. The high acoustic similarity is a necessary precondition for the system's intended use, but it is not sufficient to establish emotional comfort or engagement, which would require direct human-participant studies.

\subsection{Comparison with Existing Approaches}

Table~\ref{tab:comparison} compares the KR A1 prototype with representative existing approaches across dimensions of focus area, technology, target population, key contributions, and limitations.

\begin{table}[htbp]
\centering
\caption{Comparison of the proposed KR A1 system with existing approaches.}
\label{tab:comparison}
\small
\begin{tabular}{@{}p{1.6cm}p{2.2cm}p{2.8cm}p{2.8cm}p{2.8cm}@{}}
\toprule
\textbf{Study} & \textbf{Focus area} & \textbf{Technology employed} & \textbf{Key contributions} & \textbf{Limitations} \\
\midrule
Proposed system & Integrated humanoid interaction architecture for personalized mental-health dialogue & OpenVoice v2, Qwen2.5-3B-Instruct (Ollama), InsightFace, FAISS, Flask dashboard & System-level integration of voice cloning, emotion-aware dialogue, and local deployment; preliminary acoustic evaluation of cloned speech & No formal emotion-recognition accuracy evaluation; no clinical or HRI validation; single-speaker voice-cloning sample \\
\addlinespace
Kim et al. [39] & Social support during COVID-19 isolation using voice-cloned robots & Socially assistive robots (SAR), voice cloning & Demonstrated increased perceived social connectedness using robots with cloned voices of loved ones & Focuses solely on social connectedness; no real-time emotional adaptation or therapy personalization \\
\addlinespace
Dewangga et al. [31] & Speech synthesis for patients with speech disorders & Zero-shot TTS, speaker embedding, DNN & Improved zero-shot voice cloning for dysphonic speech clarity and naturalness & No focus on mental health or interactive therapy \\
\addlinespace
Kar [24] & General applications of robotics in healthcare & Surgical robots, rehabilitation robots & Review of robotics applications including surgery, telemedicine, and patient assistance & No specific focus on mental health or personalized therapy \\
\addlinespace
Alhussein et al. [52] & Emotion recognition from conversational speech & TCNN, affect dynamics & Novel speech-based system for detecting group emotional climate in conversations & Focuses on group-level emotions; not personalized real-time therapy \\
\addlinespace
Zhou et al. [54] & Application of NLP in smart healthcare & Statistical NLP, neural NLP, rule-based NLP & Review of NLP techniques in EHR analysis, diagnostics, and patient communication & Limited to text-based processing; lacks multimodal (voice + emotion) approaches \\
\addlinespace
Liu et al. [47] & Use of LLMs in psychiatry & LLMs, transformer-based NLP & Use of LLMs for psychiatric diagnosis support and therapeutic dialogue generation & Text-only; lacks voice and emotional adaptation capabilities \\
\addlinespace
Lin et al. [53] & Psychosocial support for older adults via robots & Telepresence robots, social robots & Demonstrated effectiveness of robot-mediated therapy in reducing isolation and anxiety & Focused on elderly care; lacks advanced emotion recognition and personalized speech synthesis \\
\bottomrule
\end{tabular}
\end{table}

The KR A1 prototype distinguishes itself from existing approaches through its system-level integration of personalized voice synthesis, contextual dialogue processing, and emotion-aware response generation within a single platform designed to support local deployment. Compared with the representative approaches listed in Table~\ref{tab:comparison}, KR A1 combines acoustic speaker-identity preservation with conversational memory and prototype crisis-alert mechanisms. However, as noted throughout this paper, these design goals remain to be validated through formal human-participant studies.

\subsection{Limitations and Threats to Validity}

The primary limitation of this study is the small, informal sample size: voice-cloning evaluation was conducted on a single speaker with 20 synthesized utterances. The absence of formal statistical significance testing, confidence intervals derived from larger samples, and held-out test sets means the results should be interpreted as preliminary feasibility evidence rather than generalizable conclusions.

The emotion-aware interaction architecture has not been benchmarked against labeled emotion datasets. The two-tier pipeline (keyword heuristic + LLM inference) provides affective labels for dialogue adaptation, but its accuracy, precision, recall, and F1 score are unknown. Claims about emotion recognition in this paper are therefore architectural, not empirical.

End-to-end response latency, memory retrieval latency, and real-time interaction quality have not been systematically measured. The system is designed for interactive operation, but latency benchmarks (mean, median, 95th percentile) are reserved for future work.

No patients or external human participants were involved in this study. All testing was informal, self-conducted by the development team. Consequently, no claims about therapeutic effectiveness, user acceptance, trust, or emotional comfort can be made. The system has not been reviewed by an institutional review board (IRB) or ethics committee, and no clinical deployment has occurred.

The voice-cloning baseline comparison, while demonstrating metric discrimination, used only one additional speaker. A rigorous speaker-verification evaluation would require a larger cohort, ROC/DET curve analysis, and reporting of equal error rate (EER) and area under the ROC curve (AUC).

\section{Conclusion and Future Work}\label{sec:conclusion}

This paper presented the design and preliminary evaluation of KR A1, a humanoid-robot interaction prototype that integrates personalized voice synthesis via OpenVoice v2, contextual dialogue processing via a locally hosted Qwen2.5-3B-Instruct model, and a two-tier emotion-aware interaction architecture. The system-level contribution lies in the integration of these components into a single platform designed to support local deployment with user recognition, persistent memory, and prototype crisis-alert and monitoring mechanisms.

The experimental evaluation, limited to the voice-cloning component, demonstrated high acoustic similarity between cloned and reference speech across spectral, prosodic, and embedding-based metrics. A same-speaker versus cross-speaker baseline comparison provided preliminary evidence that the speaker-embedding metric discriminates speaker identity within this evaluation setup, with same-speaker cosine similarity of 0.91 substantially exceeding cross-speaker (approximately 0.31) and cross-clone (approximately 0.29) conditions. These results support the technical feasibility of personalized voice synthesis within the proposed architecture.

However, the emotion-recognition accuracy, end-to-end response latency, and therapeutic or clinical impact of the integrated system have not been formally evaluated. All testing was informal, self-conducted by the development team; no patients or external human participants were involved, and the system has not been validated or deployed in any clinical setting. The therapeutic value of the proposed system remains hypothetical until evaluated in a properly designed study.

Future work should prioritize the following directions: (1)~\textit{Controlled emotion-recognition evaluation.} Benchmark the two-tier emotion pipeline against labeled emotion datasets (e.g., RAVDESS, IEMOCAP) to report accuracy, precision, recall, macro-F1, and confusion matrices. Replace the LLM-based classifier with a dedicated speech emotion recognition model (e.g., WavLM-based SER) for audio input, fused with a text-based sentiment model for multimodal inference. (2)~\textit{End-to-end latency benchmarking.} Systematically measure the latency of each pipeline stage---speech-to-text, emotion detection, memory retrieval, LLM generation, and voice synthesis---to report mean, median, 95th percentile, and worst-case response times. (3)~\textit{Speaker-verification rigor.} Expand the voice-cloning evaluation to multiple speakers and report ROC curves, AUC, and EER for the speaker-embedding metric against a larger cohort of reference and unrelated voices. (4)~\textit{Human-participant pilot study.} Conduct a small-scale pilot under appropriate IRB/ethics oversight and informed consent, assessing perceived speaker similarity, naturalness, emotional appropriateness, trust, and user comfort. A qualified mental health professional should be involved in study design and oversight. (5)~\textit{Ethical and clinical risk assessment.} Undertake a thorough analysis of the risks associated with cloning the voice of a familiar person for therapeutic use, including voice-owner consent procedures, potential for emotional dependency, perceived deception, and psychological impact of ongoing artificial relationships modeled on real persons.

\section*{Declarations}

\subsection*{Ethics Statement}

This study did not involve patients, external human participants, or a clinical trial. The system described is a technical prototype; the interaction examples shown in Table~\ref{tab:dialogue} are illustrative, constructed examples and not transcripts from real patient sessions. All hands-on testing of the system to date (including voice-cloning tests and dialogue examples) was informal, self-conducted testing carried out by a member of the development team on himself, with his full knowledge and consent; no external or vulnerable individuals were involved.

We recognize that the proposed application---cloning the voice of a person familiar to the user, including a deceased or unavailable loved one, for use in a mental-health context---raises substantive ethical questions not resolved by this feasibility study. These include the risk of fostering unhealthy emotional dependency, the potential for the interaction to be experienced as deceptive, the psychological impact of an ongoing artificial relationship modeled on a real person, and the need for informed consent both from the user and from the person whose voice is cloned (or their estate, where applicable). The system also includes a crisis-keyword detection feature intended to escalate concerning statements to a human contact; this feature has not been clinically validated and must not be relied upon as a safety mechanism in its current form.

Before this system, or any derivative of it, is used with real patients or other vulnerable individuals, we consider it essential that: (a) the study be reviewed and approved by an appropriate institutional review board or ethics committee; (b) informed consent be obtained from all participants; and (c) a qualified mental health professional be involved in the design and oversight of any deployment. No such review has yet been sought, as no such deployment has yet taken place.

\subsection*{Consent to Participate}

Not applicable. No human participants were involved in this study.

\subsection*{Data Availability Statement}

The voice recordings and reference/cloned-voice pairs analysed in this study were collected informally during development testing and are not publicly available, as they include identifiable voice data of a member of the development team. The source code for the prototype system is proprietary to AK Robotics and is not publicly available at this time. However, evaluation scripts, feature-extraction code, metric calculation code, and anonymized numerical results may be made available upon reasonable request to the corresponding author.

\subsection*{Funding Statement}

The authors declare that no external funding was received for this research. The work was self-funded by AK Robotics.

\subsection*{Competing Interests}

The authors are affiliated with AK Robotics, the company developing the system described in this manuscript, and hold leadership roles (CEO, Co-Founder, CTO) within it; authors share a family relationship. The authors declare these as competing interests. No other competing interests are declared.

\subsection*{Author Contributions}

Conceptualization: M.A., O.A., A.A. Methodology: M.A., O.A., A.A. Software: M.A., A.A. Validation: M.A., O.A. Formal analysis: M.A., O.A. Investigation: M.A., O.A., A.A. Resources: M.A., O.A, A.A. Data curation: M.A., O.A. Writing---original draft: M.A., O.A., A.A. Writing---review and editing: A.A. Visualization: M.A., O.A. Supervision: O.A. Project administration: A.A. All authors have read and agreed to the published version of the manuscript.

\subsection*{Acknowledgements}

The authors would like to express their deepest gratitude to Aanthu Kanjibhai Ramjibhai, whose visionary ideas and unwavering support have been the driving force behind this project. His inspiration and belief in the potential of technology to improve lives have been instrumental in shaping the direction of this research. The authors are also deeply thankful to Aanthu Arvindbhai Kanjibhai for his insightful ideas and constant encouragement. His support has been crucial in developing the conceptual framework for this research. The authors would also like to extend their heartfelt thanks to Aanthu Om Arvindbhai and Aanthu Mahek Arvindbhai for their invaluable contributions to coding, testing, and documentation.


\begin{thebibliography}{61}

\bibitem{ref1} Govers FX, Namuduri K. Artificial Intelligence for Robotics: Build Intelligent Robots Using ROS 2, Python, OpenCV, and AI/ML Techniques for Real-World Tasks. Birmingham: Packt Publishing; 2024.

\bibitem{ref2} Neupane S, et al. Security considerations in AI-robotics: a survey of current methods, challenges, and opportunities. IEEE Access. 2024;12:22072--22097. \url{https://doi.org/10.1109/ACCESS.2024.3363657}

\bibitem{ref3} Afrin S, Roksana S, Akram R. AI-enhanced robotic process automation: a review of intelligent automation innovations. IEEE Access. 2025;13:173--197. \url{https://doi.org/10.1109/ACCESS.2024.3513279}

\bibitem{ref4} Hans S, Chohan JS, Kumar R, Pareek S. Artificial intelligence with robotics---a bibliometric study. In: Proceedings of the 2023 10th International Conference on Computing for Sustainable Global Development (INDIACom). New Delhi, India; 2023. p. 1037--1040.

\bibitem{ref5} Kwiecie\'{n} J. Selected topics of artificial intelligence in robotics. In: Proceedings of the 2022 7th International Conference on Mechanical Engineering and Robotics Research (ICMERR). Krakow, Poland; 2022. p. 40--44. \url{https://doi.org/10.1109/ICMERR56497.2022.10097788}

\bibitem{ref6} Gupta P, et al. A comprehensive study of AI and robotics in a rapidly changing world. In: Proceedings of the 2024 International Conference on Innovations and Challenges in Emerging Technologies (ICICET). Nagpur, India; 2024. p. 1--6. \url{https://doi.org/10.1109/ICICET59348.2024.10616288}

\bibitem{ref7} Putra RVW, Marchisio A, Zayer F, Dias J, Shafique M. Embodied neuromorphic artificial intelligence for robotics: perspectives, challenges, and research development stack. In: Proceedings of the 2024 18th International Conference on Control, Automation, Robotics and Vision (ICARCV). Dubai, United Arab Emirates; 2024. p. 612--619. \url{https://doi.org/10.1109/ICARCV63323.2024.10821619}

\bibitem{ref8} Mennella C, Maniscalco U, Masala GL, Esposito M. Editorial: artificial intelligence and robotic applications for smart monitoring and assistance in healthcare services. Frontiers in Robotics and AI. 2024;11. \url{https://doi.org/10.3389/frobt.2024.1527773}

\bibitem{ref9} Mohan M, Kuchenbecker KJ. A design tool for therapeutic social-physical human-robot interactions. In: Proceedings of the 2019 14th ACM/IEEE International Conference on Human-Robot Interaction (HRI). Daegu, Korea (South); 2019. p. 727--729. \url{https://doi.org/10.1109/HRI.2019.8673202}

\bibitem{ref10} Nertinger S, Naceri A, Haddadin S. Identifying requirements for the implementation of robot-assisted physical therapy in humanoids: a user-centered design approach. In: Proceedings of the 2023 32nd IEEE International Conference on Robot and Human Interactive Communication (RO-MAN). Busan, Korea, Republic of; 2023. p. 2316--2321. \url{https://doi.org/10.1109/RO-MAN57019.2023.10309570}

\bibitem{ref11} Amato F, Di Gregorio M, Monaco C, Sebillo M, Tortora G, Vitiello G. Socially assistive robotics combined with artificial intelligence for ADHD. In: Proceedings of the 2021 IEEE 18th Annual Consumer Communications \& Networking Conference (CCNC). Las Vegas, NV, USA; 2021. p. 1--6. \url{https://doi.org/10.1109/CCNC49032.2021.9369633}

\bibitem{ref12} Reeg\aa{}rd K, Eitrheim M, Kaarstad M, Fernandes A, S\o{}rensen L. Beyond acceptance: patients perspectives on humanoid assistive robots in healthcare. In: Proceedings of the 2024 33rd IEEE International Conference on Robot and Human Interactive Communication (ROMAN). Pasadena, CA, USA; 2024. p. 851--856. \url{https://doi.org/10.1109/RO-MAN60168.2024.10731409}

\bibitem{ref13} Pillai D, Abdulaziz N. Development and evaluation of Nao humanoid robot for rehabilitation therapy. In: Proceedings of the 2024 7th International Conference on Signal Processing and Information Security (ICSPIS). Dubai, United Arab Emirates; 2024. p. 1--6. \url{https://doi.org/10.1109/ICSPIS63676.2024.10812628}

\bibitem{ref14} Elsayed N, ElSayed Z, Asadizanjani N, Ozer M, Abdelgawad A, Bayoumi M. Speech emotion recognition using supervised deep recurrent system for mental health monitoring. In: Proceedings of the 2022 IEEE 8th World Forum on Internet of Things (WF-IoT). Yokohama, Japan; 2022. p. 1--6. \url{https://doi.org/10.1109/WF-IoT54382.2022.10152117}

\bibitem{ref15} Ahuja S, Shabani A. Affective computing for social companion robots using fine-grained speech emotion recognition. In: Proceedings of the 2023 IEEE Conference on Artificial Intelligence (CAI). Santa Clara, CA, USA; 2023. p. 331--332. \url{https://doi.org/10.1109/CAI54212.2023.00146}

\bibitem{ref16} Elsayed H, Tawfik NS, Shalash O, Ismail O. Enhancing human emotion classification in human-robot interaction. In: Proceedings of the 2024 International Conference on Machine Intelligence and Smart Innovation (ICMISI). Alexandria, Egypt; 2024. p. 1--6. \url{https://doi.org/10.1109/ICMISI61517.2024.10580152}

\bibitem{ref17} Gamboa A, Dongo I, Aguilera A, Begazo R. Towards speech emotion recognition applied to social robots. In: Proceedings of the 2024 L Latin American Computer Conference (CLEI). Buenos Aires, Argentina; 2024. p. 1--10. \url{https://doi.org/10.1109/CLEI64178.2024.10700306}

\bibitem{ref18} Shenoy S, Jiang Y, Lynch T, Manuel LI, Doryab A. A self learning system for emotion awareness and adaptation in humanoid robots. In: Proceedings of the 2022 31st IEEE International Conference on Robot and Human Interactive Communication (RO-MAN). Napoli, Italy; 2022. p. 912--919. \url{https://doi.org/10.1109/RO-MAN53752.2022.9900581}

\bibitem{ref19} Alemi M, Meghdari A, Ghanbarzadeh A, Moghadam LJ, Ghanbarzadeh A. Effect of utilizing a humanoid robot as a therapy-assistant in reducing anger, anxiety, and depression. In: Proceedings of the 2014 Second RSI/ISM International Conference on Robotics and Mechatronics (ICRoM). Tehran, Iran; 2014. p. 748--753. \url{https://doi.org/10.1109/ICRoM.2014.6990993}

\bibitem{ref20} Ompico CDS, Subido ED, Banayo JN, Munsayac FET, Baldovino RG, Bugtai NT. Human movement imitation for robot therapy using the NAO robot. In: Proceedings of the 2021 IEEE 13th International Conference on Humanoid, Nanotechnology, Information Technology, Communication and Control, Environment, and Management (HNICEM). Manila, Philippines; 2021. p. 1--5. \url{https://doi.org/10.1109/HNICEM54116.2021.9731985}

\bibitem{ref21} Jeong S, et al. A robotic positive psychology coach to improve college students' wellbeing. In: Proceedings of the 2020 29th IEEE International Conference on Robot and Human Interactive Communication (RO-MAN). Naples, Italy; 2020. p. 187--194. \url{https://doi.org/10.1109/RO-MAN47096.2020.9223588}

\bibitem{ref22} Minhaj R, Hung K, Man GM-T. Development of a BCI system for enhancing human-robot interaction in cognitive stimulation therapy. In: Proceedings of the 2023 8th International Conference on Instrumentation, Control, and Automation (ICA). Jakarta, Indonesia; 2023. p. 53--58. \url{https://doi.org/10.1109/ICA58538.2023.10273122}

\bibitem{ref23} Axelsson M, Bodala IP, Gunes H. Participatory design of a robotic mental well-being coach. In: Proceedings of the 2021 30th IEEE International Conference on Robot \& Human Interactive Communication (RO-MAN). Vancouver, BC, Canada; 2021. p. 1081--1088. \url{https://doi.org/10.1109/RO-MAN50785.2021.9515356}

\bibitem{ref24} Kar S. Robotics in healthcare. In: Proceedings of the 2019 Power and Energy in International Conference (PEEIC). 2019. p. 78--83. \url{https://doi.org/10.1109/PEEIC47157.2019.8976668}

\bibitem{ref25} Tikhanoff V, Cangelosi A, Metta G. Integration of speech and action in humanoid robots: iCub simulation experiments. IEEE Transactions on Autonomous Mental Development. 2011;3(1):17--29. \url{https://doi.org/10.1109/TAMD.2010.2100390}

\bibitem{ref26} Tian Y, Li J, Lee T. Creating personalized synthetic voices from articulation impaired speech using augmented reconstruction loss. In: Proceedings of the ICASSP 2024 - 2024 IEEE International Conference on Acoustics, Speech and Signal Processing (ICASSP). Seoul, Korea, Republic of; 2024. p. 11501--11505. \url{https://doi.org/10.1109/ICASSP48485.2024.10446886}

\bibitem{ref27} Choi H-Y, Lee S-H, Lee S-W. Personalized and controllable voice style transfer with speech diffusion transformer. IEEE Transactions on Audio, Speech and Language Processing. 2025. \url{https://doi.org/10.1109/TASLPRO.2025.3533362}

\bibitem{ref28} Chen B, Du C, Yu K. Neural fusion for voice cloning. IEEE/ACM Transactions on Audio, Speech, and Language Processing. 2022;30:1993--2001. \url{https://doi.org/10.1109/TASLP.2022.3171971}

\bibitem{ref29} Huang S-F, Chen C-P, Chen Z-S, Tsai Y-P, Lee H-Y. Personalized lightweight text-to-speech: voice cloning with adaptive structured pruning. In: Proceedings of the ICASSP 2023 - 2023 IEEE International Conference on Acoustics, Speech and Signal Processing (ICASSP). Rhodes Island, Greece; 2023. p. 1--5. \url{https://doi.org/10.1109/ICASSP49357.2023.10097178}

\bibitem{ref30} Kadam S, Jikamade A, Mattoo P, Hole V. ReVoice: a neural network based voice cloning system. In: Proceedings of the 2024 IEEE 9th International Conference for Convergence in Technology (I2CT). Pune, India; 2024. p. 1--6. \url{https://doi.org/10.1109/I2CT61223.2024.10543448}

\bibitem{ref31} Dewangga D, Lestari D, Purwarianti A, Tanaya D, Azizah K, Sakti S. An evaluation of neural vocoder-based voice cloning system for dysphonia speech disorder. In: Proceedings of the 2024 27th Conference of the Oriental COCOSDA International Committee for the Co-ordination and Standardisation of Speech Databases and Assessment Techniques (O-COCOSDA). Hsinchu City, Taiwan; 2024. p. 1--7. \url{https://doi.org/10.1109/O-COCOSDA64382.2024.10800714}

\bibitem{ref32} Dinakar R, Omkar A, Bhat KK, MK NM, Hussain PA. Multispeaker and multilingual zero shot voice cloning and voice conversion. In: Proceedings of the 2023 3rd International Conference on Pervasive Computing and Social Networking (ICPCSN). Salem, India; 2023. p. 1661--1665. \url{https://doi.org/10.1109/ICPCSN58827.2023.00278}

\bibitem{ref33} Azizah K. Zero-shot voice cloning text-to-speech for dysphonia disorder speakers. IEEE Access. 2024;12:63528--63547. \url{https://doi.org/10.1109/ACCESS.2024.3396377}

\bibitem{ref34} Ortega JHJC, et al. Isolation: an interactive animated trilogy short film about mental health awareness during the COVID-19 pandemic. In: Proceedings of the 2022 IEEE 14th International Conference on Humanoid, Nanotechnology, Information Technology, Communication and Control, Environment, and Management (HNICEM). Boracay Island, Philippines; 2022. p. 1--6. \url{https://doi.org/10.1109/HNICEM57413.2022.10109398}

\bibitem{ref35} Mishra H, Choudhury GR, Pandita D. Transitioning to e-health platforms for mental wellness: the need for data-driven contemporary solutions for professionals. In: Proceedings of the 2024 International Seminar on Application for Technology of Information and Communication (iSemantic). Semarang, Indonesia; 2024. p. 89--94. \url{https://doi.org/10.1109/iSemantic63362.2024.10762698}

\bibitem{ref36} Paz JAM, Ibarra LER, Hurtado JJM. Development of technological tools in response to primary mental health care needs arising from the emergency caused by COVID-19 in Cauca. In: Proceedings of the 2023 VI Congreso Internacional en Inteligencia Ambiental, Ingenier\'{i}a de Software y Salud Electr\'{o}nica y M\'{o}vil (AmITIC). Cali, Colombia; 2023. p. 1--8. \url{https://doi.org/10.1109/AmITIC60194.2023.10366351}

\bibitem{ref37} Ahmed A, Dixit P, Khan MM. Development of an online mental well-being mobile application for Covid-19 pandemic. In: Proceedings of the 2022 6th International Conference on Computing Methodologies and Communication (ICCMC). Erode, India; 2022. p. 1553--1558. \url{https://doi.org/10.1109/ICCMC53470.2022.9754112}

\bibitem{ref38} Hamdoun S, Monteleone R, Bookman T, Michael K. AI-based and digital mental health apps: balancing need and risk. IEEE Technology and Society Magazine. 2023;42(1):25--36. \url{https://doi.org/10.1109/MTS.2023.3241309}

\bibitem{ref39} Kim JS, Shin S, Kang D, Lim Y, Kwak SS. Connecting without reaching: how voice-cloned robot can enhance mental health of isolated people during a pandemic. In: Proceedings of the 2023 32nd IEEE International Conference on Robot and Human Interactive Communication (RO-MAN). Busan, Korea, Republic of; 2023. p. 903--907. \url{https://doi.org/10.1109/RO-MAN57019.2023.10309388}

\bibitem{ref40} Kamoji S, Rozario S, Almeida S, Patil S, Patankar S, Pendhari H. Mental health prediction using machine learning models and large language model. In: Proceedings of the 2024 Second International Conference on Inventive Computing and Informatics (ICICI). Bangalore, India; 2024. p. 185--190. \url{https://doi.org/10.1109/ICICI62254.2024.00040}

\bibitem{ref41} Han G, Liu W, Huang X, Borsari B. Chain-of-interaction: enhancing large language models for psychiatric behavior understanding by dyadic contexts. In: Proceedings of the 2024 IEEE 12th International Conference on Healthcare Informatics (ICHI). Orlando, FL, USA; 2024. p. 392--401. \url{https://doi.org/10.1109/ICHI61247.2024.00057}

\bibitem{ref42} Wang X, Liu K, Wang C. Knowledge-enhanced pre-training large language model for depression diagnosis and treatment. In: Proceedings of the 2023 IEEE 9th International Conference on Cloud Computing and Intelligent Systems (CCIS). Dali, China; 2023. p. 532--536. \url{https://doi.org/10.1109/CCIS59572.2023.10263217}

\bibitem{ref43} Chen K, Kang X, Lai X, Ni Z. Enhancing emotional support capabilities of large language models through cascaded neural networks. In: Proceedings of the 2023 4th International Conference on Computer, Big Data and Artificial Intelligence (ICCBD+AI). Guiyang, China; 2023. p. 318--326. \url{https://doi.org/10.1109/ICCBD-AI62252.2023.00060}

\bibitem{ref44} Yang M, Tao Y, Cai H, Hu B. Behavioral information feedback with large language models for mental disorders: perspectives and insights. IEEE Transactions on Computational Social Systems. 2024;11(3):3026--3044. \url{https://doi.org/10.1109/TCSS.2024.3397403}

\bibitem{ref45} Oniani D, et al. Enhancing large language models for clinical decision support by incorporating clinical practice guidelines. In: Proceedings of the 2024 IEEE 12th International Conference on Healthcare Informatics (ICHI). Orlando, FL, USA; 2024. p. 694--702. \url{https://doi.org/10.1109/ICHI61247.2024.00111}

\bibitem{ref46} Manoj R, G N. A comprehensive investigation on leveraging generative AI and large language models in the healthcare domain. In: Proceedings of the 2024 IEEE 12th Region 10 Humanitarian Technology Conference (R10-HTC). Kuala Lumpur, Malaysia; 2024. p. 1--6. \url{https://doi.org/10.1109/R10-HTC59322.2024.10778845}

\bibitem{ref47} Liu Z, et al. Large language models in psychiatry: current applications, limitations, and future scope. Big Data Mining and Analytics. 2024;7(4):1148--1168.

\bibitem{ref48} Riccio T. Sophia robot: an emergent ethnography. TDR: The Drama Review. 2021;65(3):42--77. \url{https://doi.org/10.1017/S1054204321000319}

\bibitem{ref49} Retto J. Sophia, first citizen robot of the world. 2017.

\bibitem{ref50} Fuchs T. Understanding Sophia? On human interaction with artificial agents. Phenomenology and the Cognitive Sciences. 2024;23:21--42. \url{https://doi.org/10.1007/s11097-022-09848-0}

\bibitem{ref51} Zhou T, Xu T. Sophia-in-audition: virtual production with a robot performer. arXiv:2402.06978. 2024.

\bibitem{ref52} Alhussein G, Alkhodari M, Khandoker AH, Hadjileontiadis LJ. Novel speech-based emotion climate recognition in peers' conversations incorporating affect dynamics and temporal convolutional neural networks. IEEE Access. 2025;13:16752--16769. \url{https://doi.org/10.1109/ACCESS.2025.3529125}

\bibitem{ref53} Lin X, et al. Mental health problems of older adults during the COVID-19 pandemic: effects, recommendations and healthcare robot-mediated therapy. In: Proceedings of the 2022 12th International Conference on CYBER Technology in Automation, Control, and Intelligent Systems (CYBER). Baishan, China; 2022. p. 1061--1065. \url{https://doi.org/10.1109/CYBER55403.2022.9907525}

\bibitem{ref54} Zhou B, Yang G, Shi Z, Ma S. Natural language processing for smart healthcare. IEEE Reviews in Biomedical Engineering. 2024;17:4--18. \url{https://doi.org/10.1109/RBME.2022.3210270}

\bibitem{ref55} Qin Z, Zhao Z, Xie Y, Wu M. OpenVoice: versatile instant voice cloning. arXiv:2312.01479. 2023. The v2 release (April 2024) was used in this work; it provides improved audio quality and native multilingual support over the original version.

\bibitem{ref56} Qwen et al. Qwen2.5 technical report. 2024. \url{https://qwenlm.github.io/blog/qwen2.5-llm/}

\bibitem{ref57} Deng J, Guo J, Xue N, Zafeiriou S. ArcFace: additive angular margin loss for deep face recognition. In: Proceedings of the IEEE/CVF Conference on Computer Vision and Pattern Recognition (CVPR). 2019. p. 4690--4699.

\bibitem{ref58} Johnson J, Douze M, J\'{e}gou H. Billion-scale similarity search with GPUs. IEEE Transactions on Big Data. 2019;7(3):535--547. \url{https://doi.org/10.1109/TBDATA.2019.2921572}

\bibitem{ref59} Reimers N, Gurevych I. Sentence-BERT: sentence embeddings using Siamese BERT-networks. In: Proceedings of the 2019 Conference on Empirical Methods in Natural Language Processing (EMNLP). Hong Kong, China; 2019. p. 3982--3992. \url{https://doi.org/10.18653/v1/D19-1410}. The specific all-mpnet-base-v2 model used in this work is available from the Sentence Transformers model repository.

\bibitem{ref60} Zhang A. SpeechRecognition [software]. GitHub; 2023. \url{https://github.com/Uberi/speech_recognition}

\bibitem{ref61} Regehr K, et al. SER evals: in-domain and out-of-domain benchmarking for speech emotion recognition. arXiv:2408.07851. 2024.

\end{thebibliography}
\end{document}